\documentclass[a4paper]{PoS}
\pdfoutput=1
\usepackage{amsmath,amscd}
\usepackage{pb-diagram}
\usepackage{booktabs}
\usepackage{braket}

\DeclareMathOperator{\Tr}{Tr}

\title{SO(2N) and SU(N) gauge theories}

\ShortTitle{SO(2N) and SU(N) gauge theories}

\author{\speaker{Richard Lau}\footnote{Funded by the Science and Technology Facilities Council.} \ and Michael Teper\\
        Rudolf Peierls Centre for Theoretical Physics, University of Oxford\\
        E-mail: \email{richard.lau@physics.ox.ac.uk} \\
 		E-mail: \email{m.teper1@physics.ox.ac.uk}}

\abstract{We present our preliminary results of $SO(2N)$ gauge theories, approaching the large-$N$ limit. 
$SO(2N)$ theories may help us to understand QCD at finite chemical potential since there is an orbifold equivalence between $SO(2N)$ and $SU(N)$ gauge theories at large-$N$ and $SO(2N)$ theories do not have the sign problem present in QCD. 
We consider the string tensions, mass spectra, and deconfinement temperatures in the $SO(2N)$ pure gauge theories in 2+1 dimensions, comparing them to their corresponding $SU(N)$ theories.}

\FullConference{31st International Symposium on Lattice Field Theory LATTICE 2013\\
                 July 29 - August 3, 2013\\
                 Mainz, Germany}

\begin{document}

\section{Introduction}

$SO(N)$ gauge theories do not have a sign problem at finite chemical potential \cite{cherman11}, unlike $SU(N)$ QCD theories, and share particular equivalences with $SU(N)$ gauge theories. 
There are, of course, direct equivalences between specific groups that share common Lie algebras, such as $SU(2) \sim SO(3)$ or $SU(4) \sim SO(6)$, and we showed in a previous paper that each pair of gauge groups share particular physical characteristics between their pure gauge theories \cite{bursa13}.

However, there is also an orbifold equivalence between $SO(2N)$ and $SU(N)$ theories \cite{cherman11}. 
Under this orbifold equivalence, we can obtain an $SU(N)$ QCD theory through a projection symmetry applied to a parent $SO(2N)$ QCD-like gauge theory. 
This equivalence holds if we take the large-$N$ limit whilst relating the couplings $g$ in the two theories by 
\begin{align}
 \left. g^{2} \right|_{SU(N \rightarrow \infty)} &=  \left. g^{2} \right|_{SO(2N \rightarrow \infty)} 
 \label{coupling}
\end{align}

This large-$N$ equivalence, together with the lack of a sign problem in $SO(N)$ gauge theories, indicates that the properties of $SO(2N)$ gauge theories may provide a potential starting point towards answering problems with $SU(N)$ QCD theories at finite chemical potential \cite{cherman11}.

In this contribution, we calculate physical quantities in $SO(2N)$ pure gauge theories. 
We know that we can extrapolate to the large-$N$ limit for both $SO(2N)$ and $SU(N)$ theories by keeping the t'Hooft coupling $g^{2}N$ constant. We also know that the leading correction between finite $N$ and the large-$N$ limit is $\mathcal{O}(1/N)$ for $SO(2N)$ and $\mathcal{O}(1/N^2)$ for $SU(N)$. By comparing these values at large-$N$ limit, we hope to relate the two gauge theories, a process summarised in \eqref{eqn:equiv}.
\begin{align}
\begin{diagram}
\dgARROWLENGTH=6em
\node{SU(N \rightarrow\infty)} \arrow{e,t,<>}{\text{large-$N$ equivalence}} \arrow{s,l,<>}{\mathcal{O}\left(\frac{1}{N^2}\right)\text{ corrections}} 
    \node{SO(2N \rightarrow\infty)} \arrow{s,r,<>}{\mathcal{O}\left(\frac{1}{N}\right)\text{ corrections}} \\
\node{SU(N)} 			\node{SO(2N)} 
\end{diagram}
\label{eqn:equiv}
\end{align}

In this contribution, we consider the string tension, mass spectra, and deconfinement temperatures in $SO(N)$ pure gauge theories for $N=$ 6, 8, 12, 16. The lattice action for an $SO(N)$ gauge theory is
\begin{align}
S=\beta\sum_{p}\left( 1-\frac{1}{N} \Tr U_p \right) \qquad\qquad \beta=\frac{2N}{ag^2} 
\end{align}
We calculate these physical quantities on several lattice spacings before extrapolating to the continuum limit for each $SO(N)$ value. Using \eqref{eqn:equiv}, we can then extrapolate to the large-$N$ limit.

Since lattice gauge theories generally have a bulk transition separating strong and weak coupling regions, we need to know where this transition is so that we can extrapolate to the continuum limit on the weak coupling side. 
Furthermore, we need this transition to occur at coupling values corresponding to lattice volumes at which we can reasonably calculate quantities. 
Otherwise, the volumes may become too large to obtain results.
This is the problem in $D=3+1$ dimensions, where the bulk transition occurs at a very small lattice spacing and so the volumes can become too large to obtain continuum extrapolations \cite{forcrand03}. 
However, in $D=2+1$ dimensions, the bulk transition occurs at larger lattice spacings and we can obtain continuum extrapolations at reasonable volumes. It is for this reason that we initially use $D=2+1$ lattices for our calculations.

In this contribution, we publish our preliminary results for these measurements. We will publish further results, including some for $D=3+1$ dimensions, in future papers.

\section{String Tensions}

We can obtain string tensions $\sigma$ by using correlators of Polyakov loop operators $l_P (t)$ to extract the mass of the lightest flux loop winding around the spatial torus. 
From a mass $m_{P}(l)$ of a Polyakov loop of lattice length $l$, we can obtain the string tension using the Nambu-Goto model \cite{athenodorou11}
\begin{align}
	m_{P}(l)=\sigma l\left(1-\frac{\pi}{3\sigma l^2}\right)^{\frac{1}{2}} 
	\label{eqn:nambugoto}
\end{align}

\begin{figure}[h] 
  	\includegraphics[width=\textwidth]{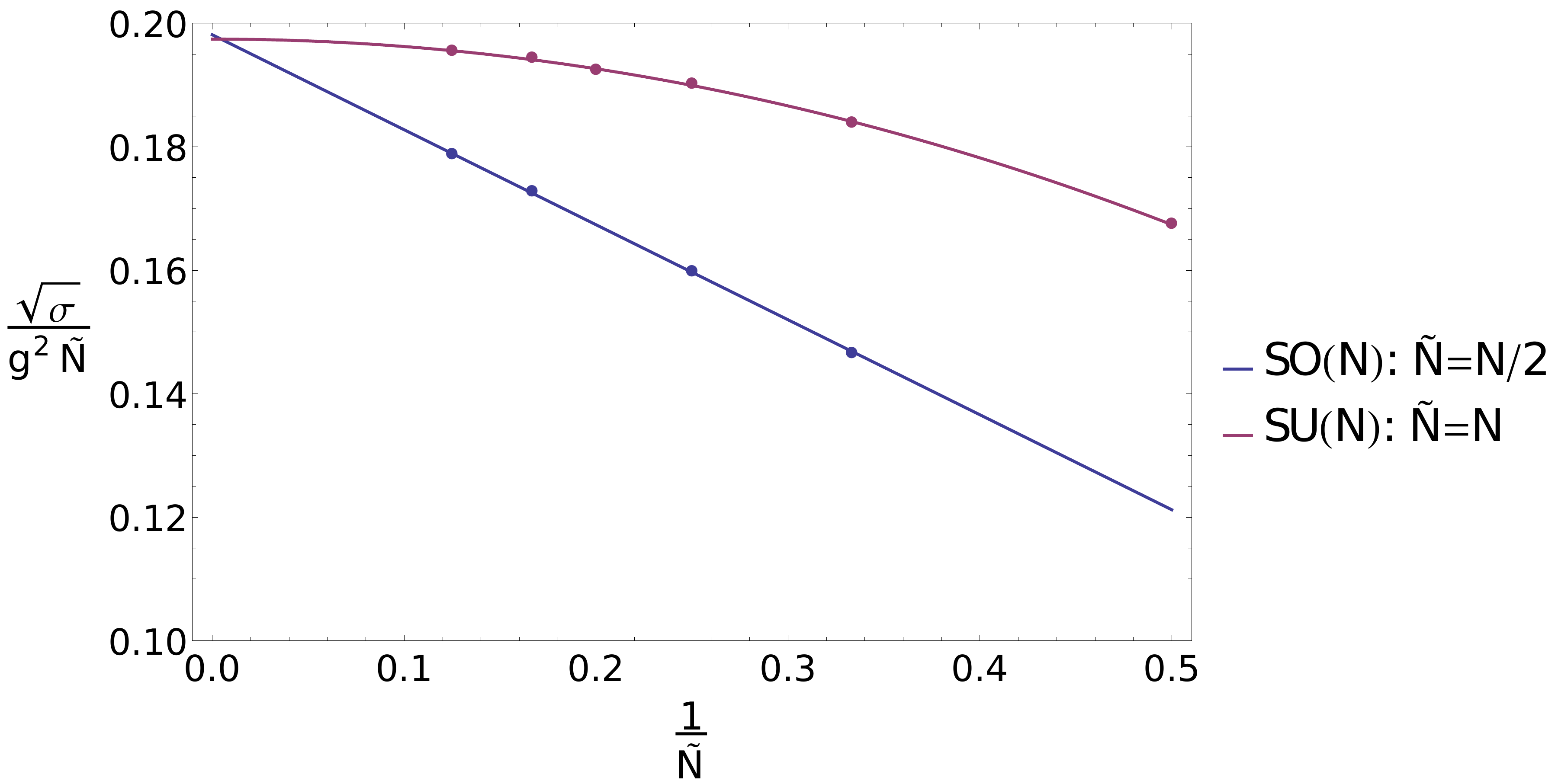} 
	\caption{Large-$N$ extrapolation of continuum string tensions.}
	\label{fig:stringtension-soinf}
\end{figure}

We obtained the continuum string tensions for $SO(N)$ for $N=$ 6, 8, 12, 16. 
We show these values in Figure~\ref{fig:stringtension-soinf}, comparing them to known values for $SU(N)$ gauge groups \cite{bringoltz07}. 
In Figure~\ref{fig:stringtension-soinf}, we rescaled $N\rightarrow \tilde{N}$ such that $\tilde{N}=N/2$ for $SO(N)$ and $\tilde{N}=N$ for $SU(N)$ to make the comparison between the two gauge theories clearer. We fitted the $SO(N)$ values with a first order fit in $1/\tilde{N}$ and the $SU(N)$ values with a first order fit in $1/\tilde{N}^2$.

In Figure~\ref{fig:stringtension-soinf}, we see that these values approach each other in the large-$N$ limit (after the appropriate rescaling). Furthermore, the values in this limit agree within errors, as shown in Table~\ref{tab:stringtension-soinf}.

\begin{table}[h]
\centering
\begin{tabular}{ |c| c| }
  \hline                        
Gauge group 	& $\left. \frac{\sqrt{\sigma}}{g^2\tilde{N}} \right|_{\tilde{N} \rightarrow \infty}$ \\ 
  \hline             
  \hline
 $SO(2N)$ 		& $0.1981(6)$  \\
  \hline                        
  $SU(N)$ 		& $0.1974(2)$ \\
  \hline                      
\end{tabular}
	\caption{Large-$N$ string tensions for $SO(2N)$ and $SU(N)$.}
	\label{tab:stringtension-soinf}  
\end{table}

\section{Mass Spectra}

\begin{figure}[h] 
  	\includegraphics[width=0.9\textwidth]{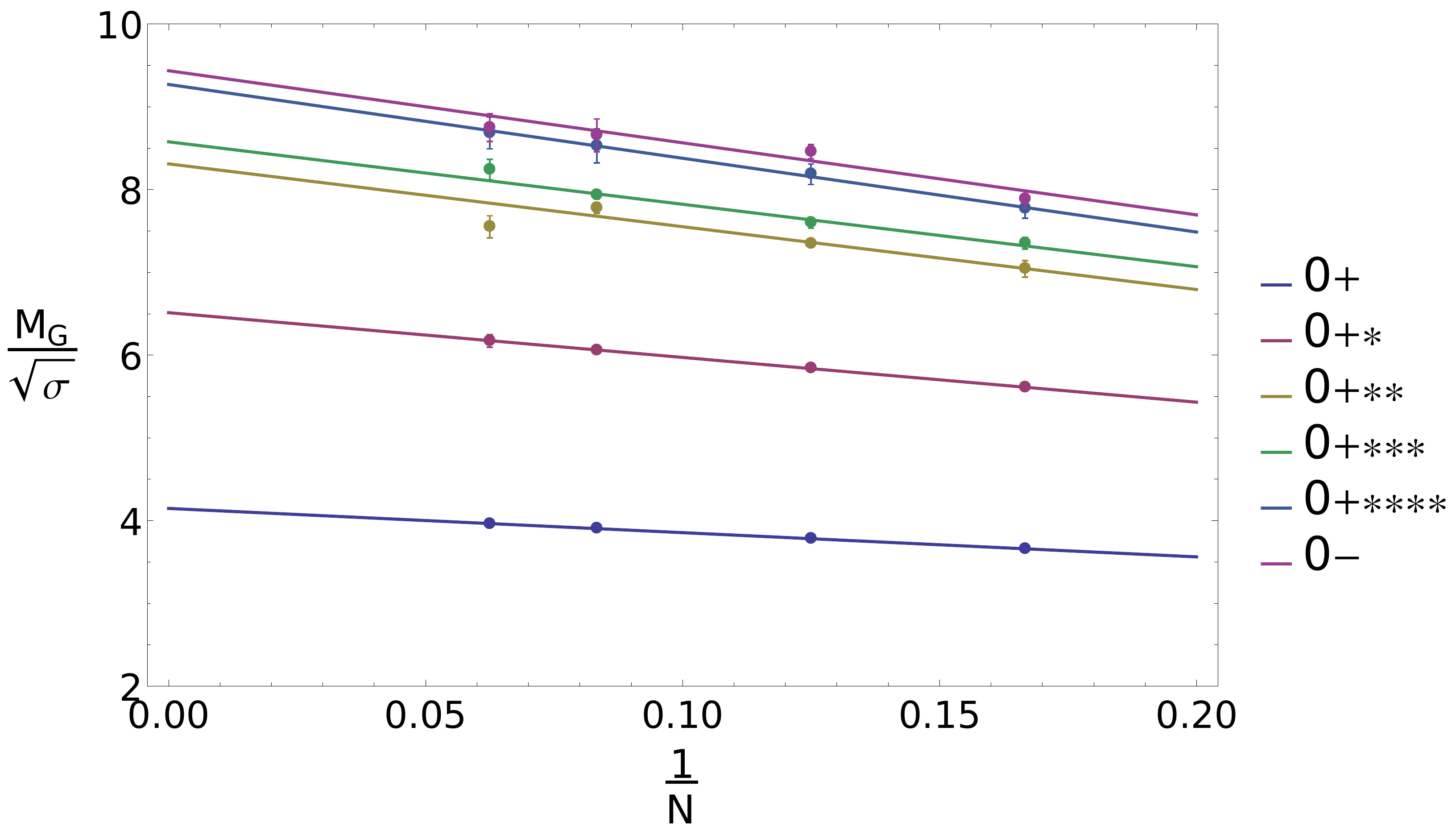} 
	\caption{Large-$N$ extrapolation of continuum $0^{+/-}$ glueball masses in $SO(N)$ gauge theories.}
	\label{fig:mass0-soinf}
\end{figure}

\begin{figure}[h] 
  	\includegraphics[width=0.9\textwidth]{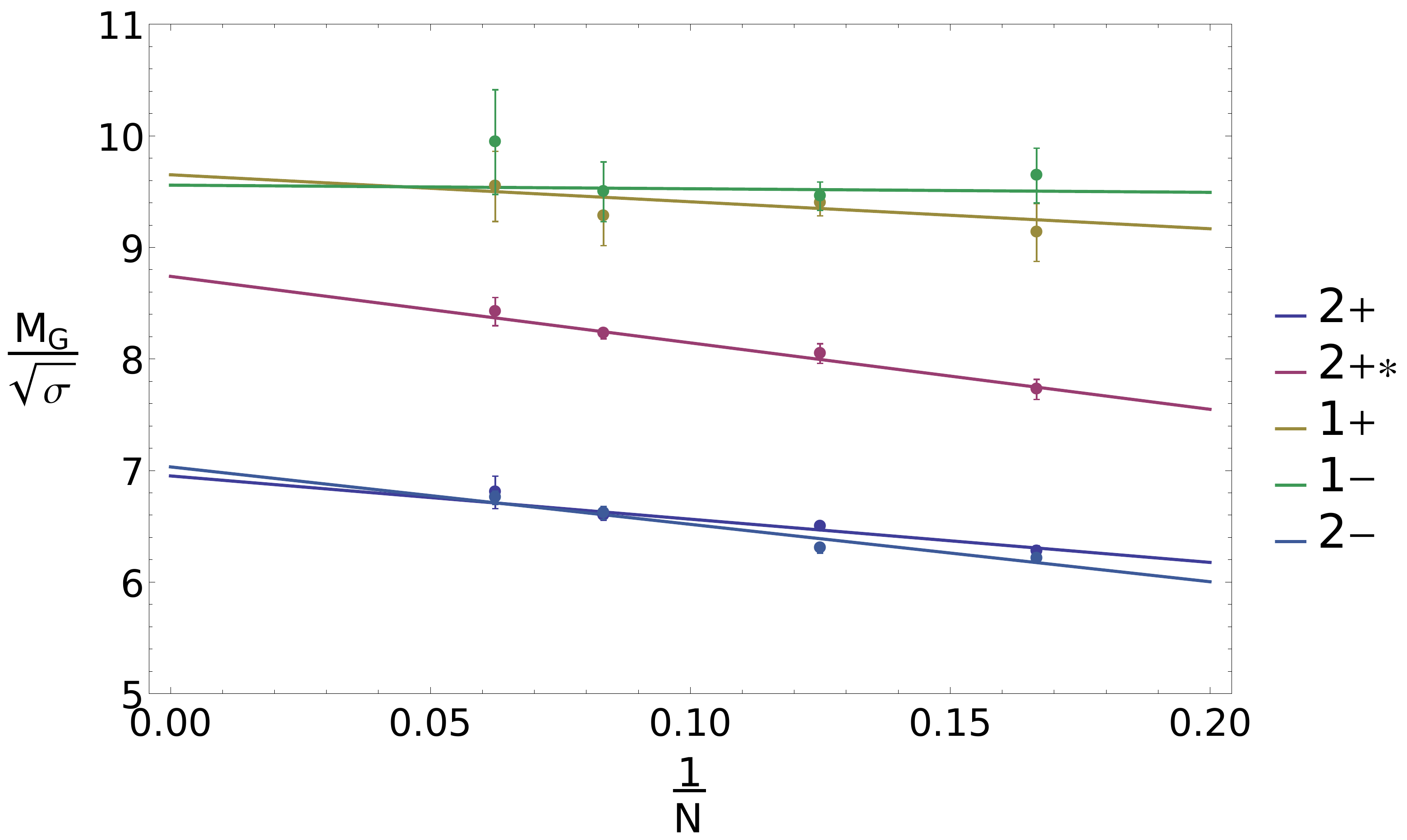} 
 	\caption{Large-$N$ extrapolation of continuum $1^{+/-}$ and $2^{+/-}$ glueball masses in $SO(N)$ gauge theories.}
	\label{fig:mass12-soinf}
\end{figure}

We can obtain mass spectra by using correlators of operators projecting on to $J^P$ glueballs with spin $J$ and parity $P=\pm$ (since $SO(N)$ traces are real, charge conjugation is necessarily positive). 
We used a variational method to construct operators that best project on to these states \cite{teper99}. 
We obtained the lightest and excited $0^+$ and $2^+$ states as well as the lightest $0^-$, $1^+$, $1^-$, and $2^-$ states for $SO(N)$ with $N=$ 6, 8, 12, 16.
Using these continuum values, we obtained the mass spectra in the large-$N$ limit, as shown in Figure~\ref{fig:mass0-soinf} and Figure~\ref{fig:mass12-soinf}. 
The fits are first order in $1/N$.

We can compare the large-$N$ values for the lightest states to the corresponding known large-$N$ values for $SU(N)$ theories \cite{lucini02}. These values, shown in Table~\ref{tab:mass-soinf}, agree within errors.

\begin{table}[h]
\centering
\begin{tabular}{ |c|c|c| }
	\hline
  $J^P $ 	& $SO(2N\rightarrow \infty)$ 	& $SU(N\rightarrow \infty)$ \\
  \hline
  \hline
  $0^{+}$ 		& 4.14(3)		& 4.11(2)		\\
  $0^{-}$ 		& 9.44(22)		& 9.02(30)		\\
  $1^{+}$		& 9.65(41) 		& 9.98(25)		\\
  $1^{-}$		& 9.56(48)		& 10.06(40)		\\
  $2^{+}$		& 6.95(9) 		& 6.88(6)		\\
  $2^{-}$		& 7.03(8) 		& 6.89(21)		\\
  \hline
\end{tabular}
 	\caption{Large-$N$ mass spectra for $SO(2N)$ and $SU(N)$.}
	\label{tab:mass-soinf}
\end{table}

\section{Deconfining Temperatures}

\begin{figure}[h]
  	\includegraphics[width=0.9\textwidth]{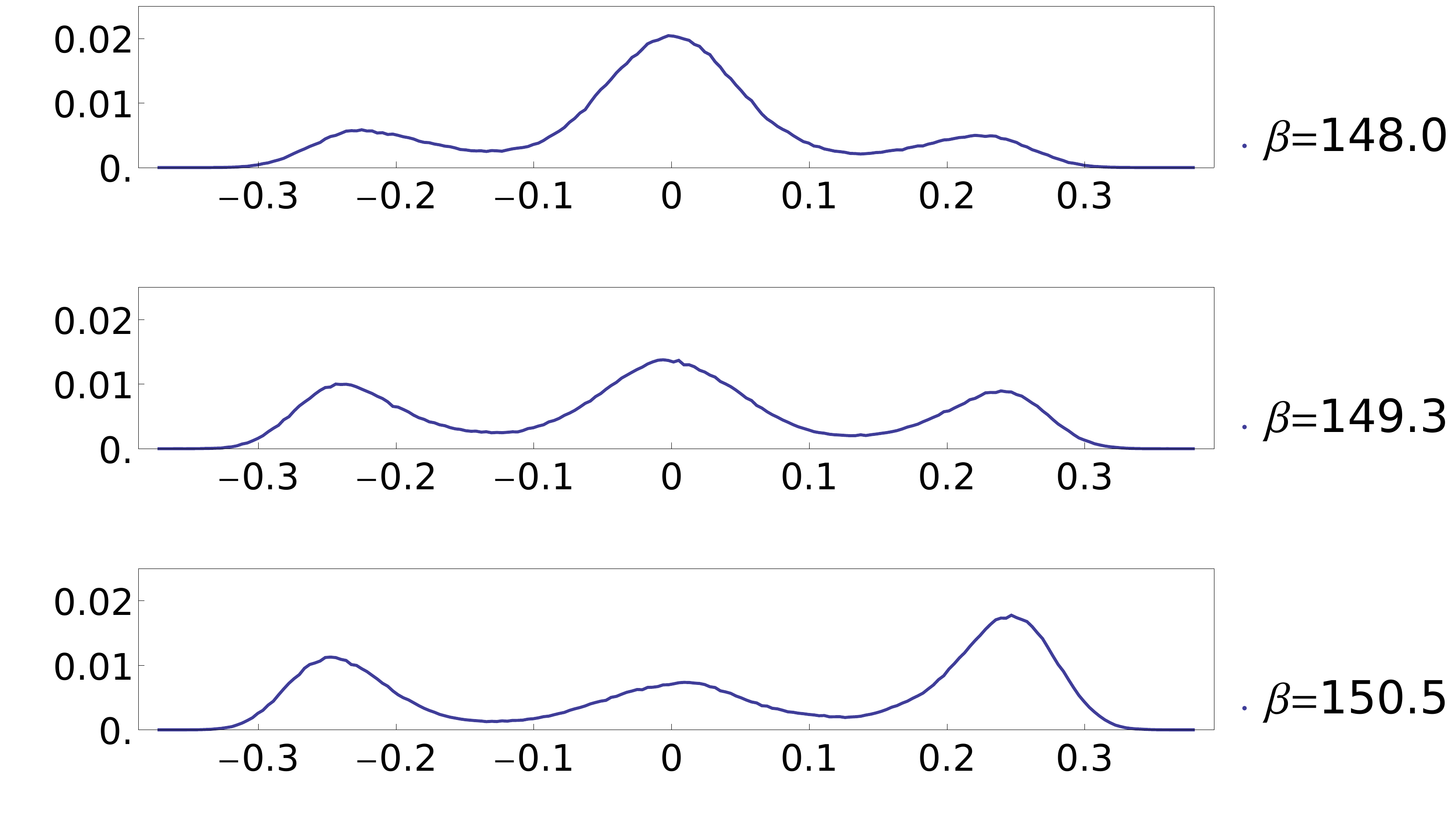} 
   	\caption{Histograms of $\braket{\overline{l_p}}$ in $SO(16)$ at several values of $\beta$ around $\beta_c$ for an $8^{2}3$ lattice.}
	\label{fig:phase-so16-3_8-tunnelling}
\end{figure}

We expect $SO(2N)$ gauge theories to deconfine at some temperature $T=T_c$, just like $SU(N)$ gauge theories.
We can search for the deconfinement temperature by using an `order parameter' $O$ such as the temporal plaquette $\overline{U_t}$ or the Polyakov loop $ \overline{l_p} $.
We can identify the range of $\beta$ in which the deconfinement phase transition occurs by examining histograms of the expectation values of the order parameters.
We show one such example of a set of histograms in Figure~\ref{fig:phase-so16-3_8-tunnelling}.
In the first histogram, we can see that the order parameter $\braket{\overline{l_p}}$ is centred around zero since the $SO(2N)$ gauge theory has a $\mathbb{Z}_2$ symmetry. 
As we approach $\beta=\beta_c$ corresponding to the value of $T_c$, this $\mathbb{Z}_2$ symmetry spontaneously breaks.
We can see this symmetry breaking in the histograms since, as we increase $\beta$ towards $\beta_c$, peaks for the deconfined phase at non-zero values of $\braket{\overline{l_p}}$ appear and grow, whilst the peak for the confined phase shrinks.

We can identify the deconfinement temperature by using susceptibilities $\chi_{_{O}} \sim \braket{O^2}-\braket{O}^2$ for an order parameter $O$.
To do this, we obtain specific values of susceptibilities around $\beta_c$.
Plots of these susceptibilities $\chi_{_{O}}$ against $\beta$ in the region around $\beta_c$ then form a peak with a maximum at $\beta_c$.
In order to calculate the susceptibility at an arbitrary value of $\beta=\beta'$ around $\beta_c$, we use reweighting methods \cite{ferrenberg89}.
We can consider the generation of lattice configurations as sampling an underlying density of states that is independent of $\beta$.
From any one run at $\beta=\beta''$, we can reconstruct the density of states in the neighbourhood of $\beta''$, and from several runs, we can evaluate the density of states extensively over a range of $\beta$.
We can then use this reconstructed density of states to obtain observables at an arbitrary value of $\beta'$ within that range.

\begin{figure}[h] 
  	\includegraphics[width=0.8\textwidth]{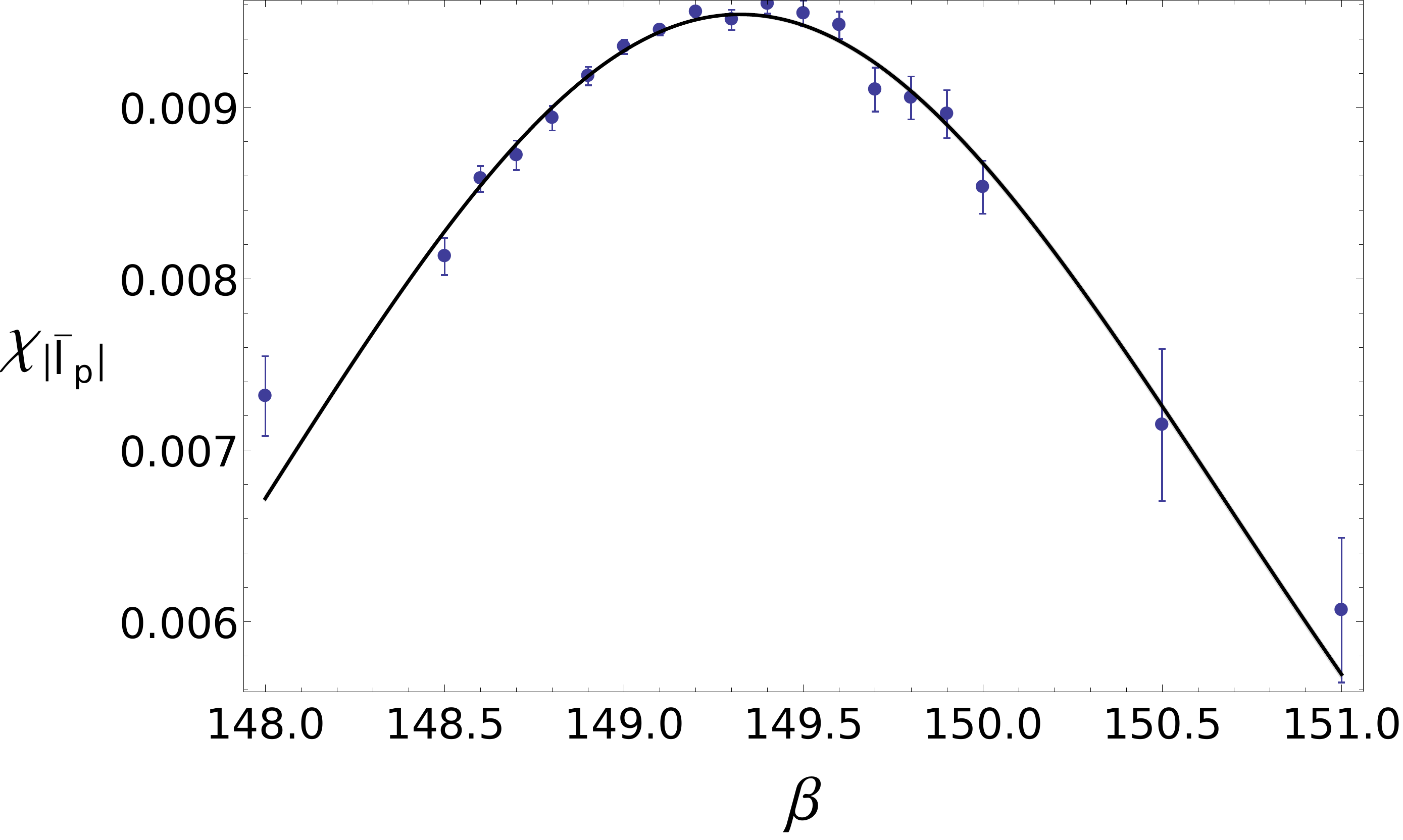} 
   	\caption{$\chi_{\lvert \overline{l_p} \rvert }$ in $SO(16)$ for an $8^{2}3$ lattice. In this example, we find that $\beta_c = 149.32(1)$.}
	\label{fig:phase-so16-3_8-lp}	
\end{figure}

We calculated the susceptibilities $\chi_{\overline{U_t}}$ the temporal plaquette $\overline{U_t}$ and $\chi_{\lvert \overline{l_p} \rvert }$ of the absolute value of the Polyakov loop $\chi_{\lvert \overline{l_p} \rvert}$ for a range of different volumes, and then reweighted the data to obtain $\beta_c$ for each volume.
We show an example of this in Figure~\ref{fig:phase-so16-3_8-lp}. Here, the points represent susceptibility values $\chi_{\lvert \overline{l_p} \rvert }$ for independent runs at specific values of $\beta$.
The curve represents the reweighted susceptibility values using the data from each independent run to reconstruct the density of states.
The value of $\beta$ with the maximum reweighted susceptibility is then the value of $\beta_c$ for this volume.

\begin{figure}[h]
  	\includegraphics[width=0.45\textwidth]{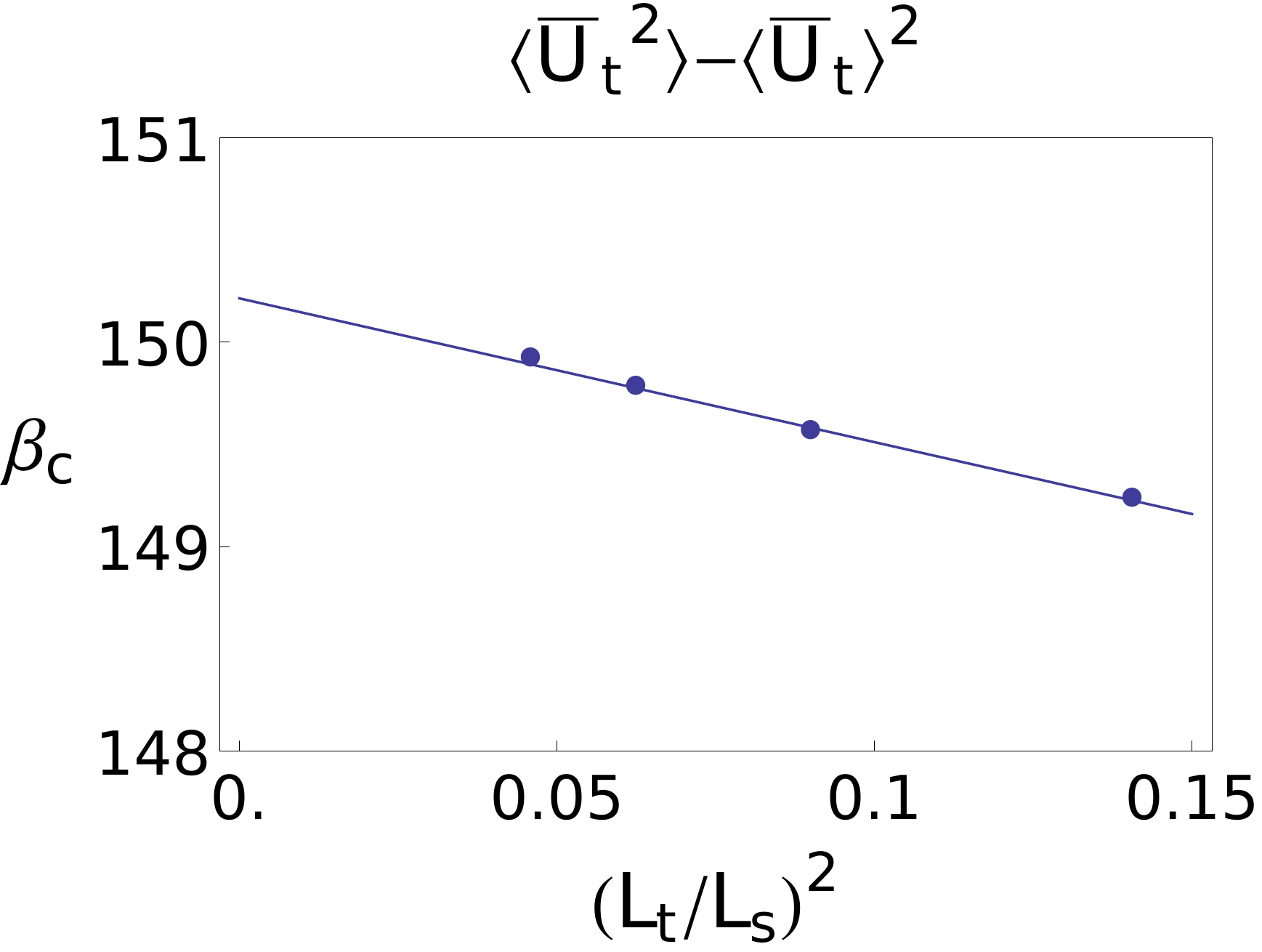} \qquad 
  	\includegraphics[width=0.45\textwidth]{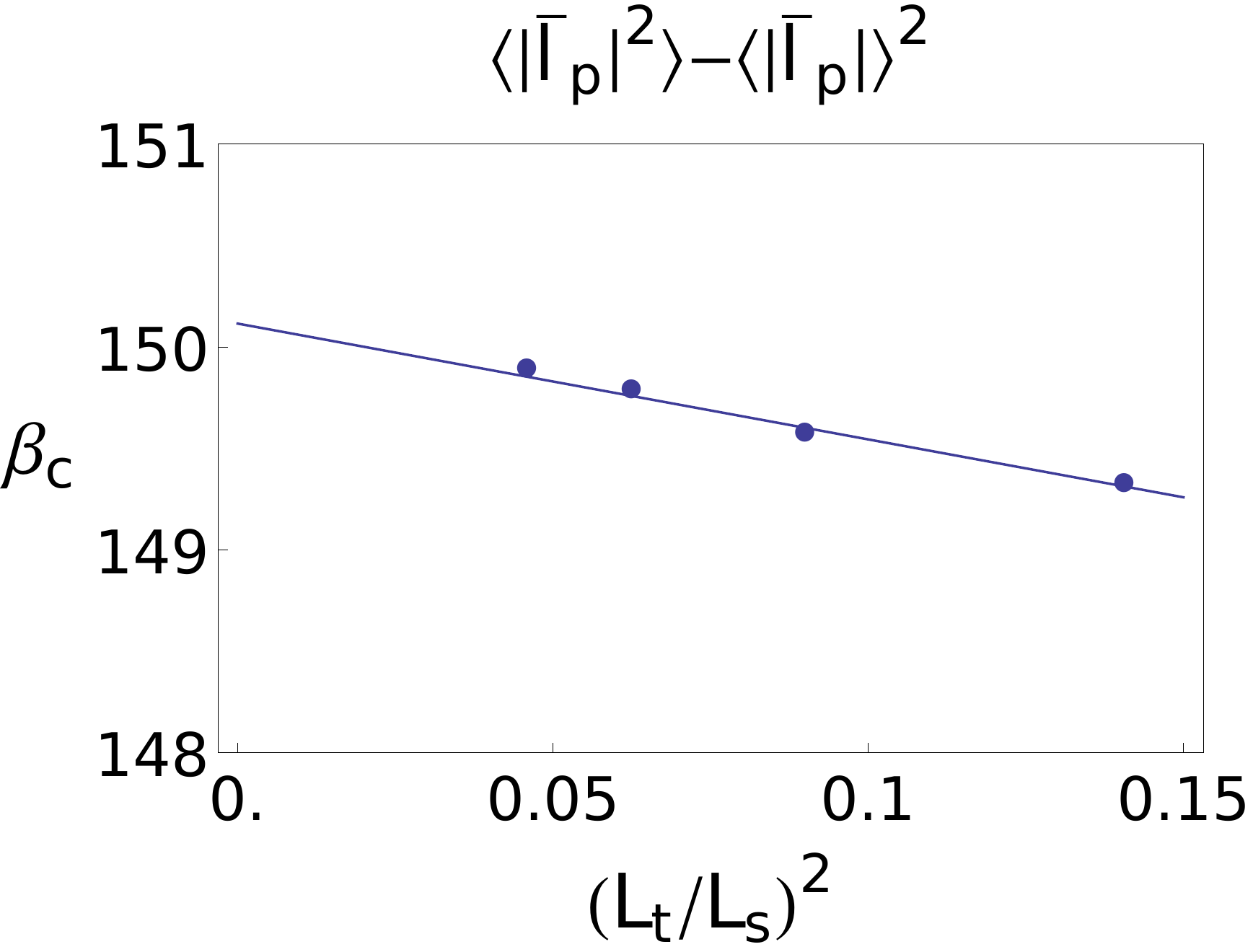} 
	\caption{$\beta_c$ in the large spatial volume limit in $SO(16)$ at $T_c=\frac{1}{3a}$ with $L_s=$ 8, 10, 12, 14. In this example, we find from $\chi_{\overline{U_t}}$ that $\beta_c(L_s\rightarrow\infty)=150.21(3)$ and from $\chi_{\lvert \overline{l_p} \rvert}$that $\beta_c(L_s\rightarrow\infty) = 150.16(2)$ .}
	\label{fig:phase-so16-lt_3}
\end{figure}

For a volume $L_s^2L_t$ with $L_s\gg L_t$, we can set the temperature $T=1/(aL_t)$. Having obtained the value of $\beta_c$ for a range of volumes, we can then extrapolate to the large spatial volume limit $L_s \rightarrow \infty$ for a fixed value of $T_c$ to find the value of $\beta_c$ in that limit. We show an example of such an extrapolation in Figure~\ref{fig:phase-so16-lt_3}. The fits are first order in $(L_t/L_s)^2$.

Having obtained $\beta_c$ in this limit, we can calculate a dimensionless quantity such as $T_c/\sqrt{\sigma}$ by evaluating observables at this value of $\beta=\beta_c$, and then extrapolate those quantities to the large-$N$ limit.
We gave our preliminary values for $T_c/\sqrt{\sigma}$ in the large-$N$ limit in \cite{bursa13}. 
We compare those values to known values for $SU(N)$ gauge theories \cite{liddle} in Table~\ref{tab:tc_soinf}, and we see that they agree within errors. 
We will publish further calculations of these deconfining temperatures in future papers.

\begin{table}[h]
\centering
\begin{tabular}{ |c|c| }
  \hline
  Gauge group 	&  $T_c/\sqrt{\sigma}$\\
  \hline\hline
  $SO(2N\rightarrow \infty)$		& 0.924(20)		 \\ 
  \hline
  $SU(N\rightarrow \infty)$ 	& 0.903(23)		 \\
  \hline
\end{tabular}
	\caption{Large-$N$ deconfinement temperatures for $SO(2N)$ and $SU(N)$.}
	\label{tab:tc_soinf}
\end{table}

\section{Conclusions}

We see that there there is evidence supporting the large-$N$ equivalence between $SO(2N)$ and $SU(N)$ gauge theories. 
In particular, we see that these pure gauge theories in $D=2+1$ dimensions have matching physical properties at large-$N$ for their string tensions, mass spectra, and deconfining temperatures. 
Following these preliminary results, we will publish further results in future papers. 
However, these preliminary results indicate that $SO(2N)$ theories may indeed provide a starting point for answering problems with $SU(N)$ QCD theories at finite chemical potential.


\end{document}